\documentclass[a4paper,12pt,openright]{article}
\usepackage{verbatim} 
\usepackage[pdftex]{graphicx} 
\usepackage{graphicx} 
\usepackage{caption} 
\usepackage{subcaption}\usepackage{amsmath} 
\usepackage{wrapfig} 
\usepackage[margin=1.0in]{geometry}
\usepackage[margin=1cm]{caption} 

\title{
  Exact solutions for silicon photomultipliers models and application to measurements
}
\author{
  {\bf Enrico Junior Schioppa}
  \\
  \emph{CERN, CH-1211 Geneva 23, Switzerland}
  \\
  enrico.junior.schioppa@cern.ch
}

\begin{document}

\maketitle
\abstract{
  Dark count rate and correlated noise rate are among the main parameters that characterize silicon photomultipliers (SiPM). Typically, these parameters are evaluated by applying approximate formulas, or by fitting specific models, to the measured SiPM noise spectra. Here a novel approach is presented, where exact formulas are derived from a statistical model of dark counts and correlated noise generation. The method allows one to measure the true value of such parameters from the areas of just the first peaks in the dark spectrum. A numerical analysis shows the accuracy of the method.
}


\section{Introduction}
\begin{figure}
  \centering
  \includegraphics[width=0.5\textwidth]{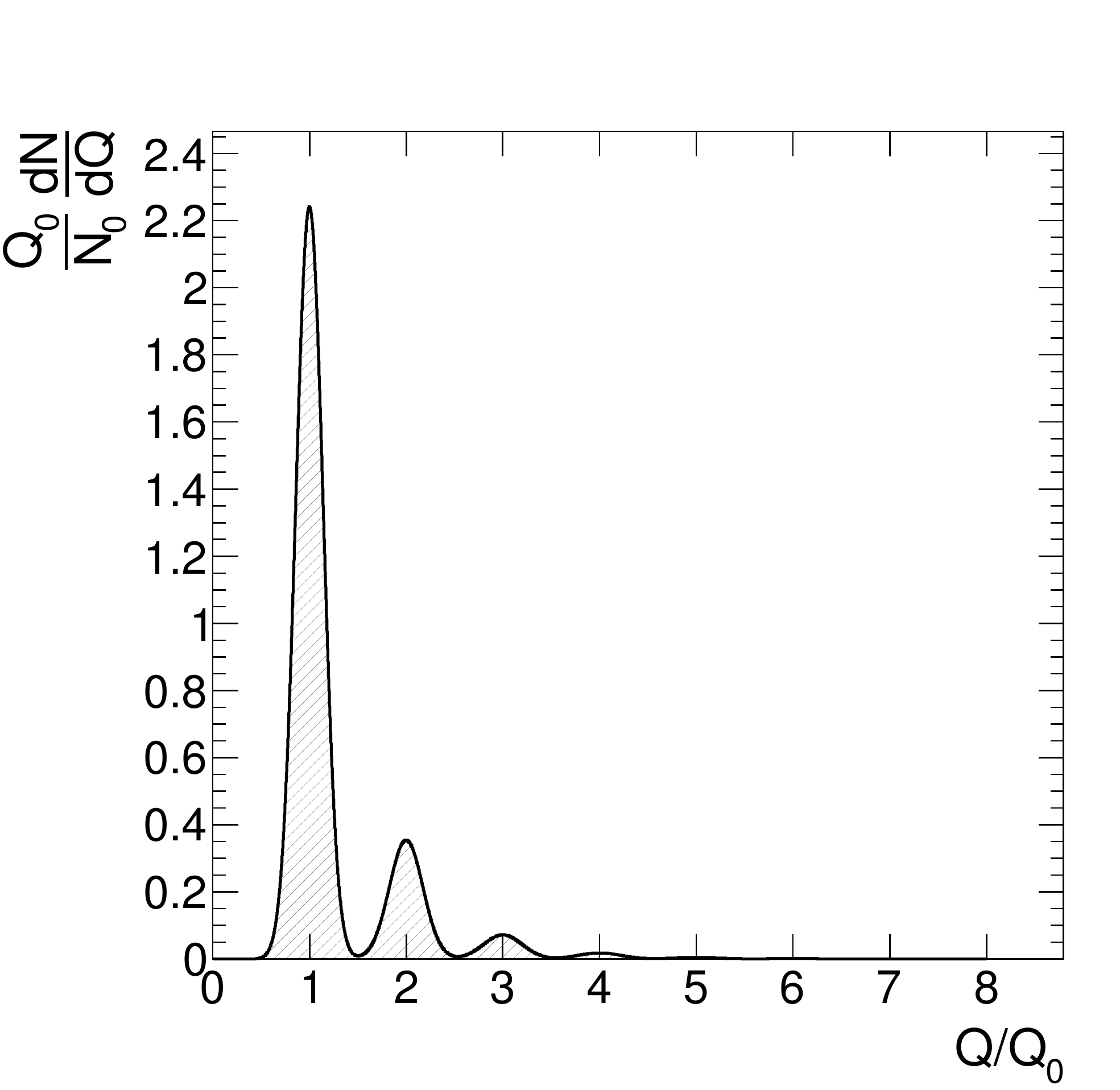}
  \caption{A typical SiPM noise charge spectrum (simulation).}
  \label{fig: SPE}
\end{figure}

Silicon photomultipliers (SiPM) are nowadays largely employed for the detection of light in a number of applications, ranging from gamma ray astronomy~\cite{FACT,SST-1M,SCT} to positron emission tomography (PET)~\cite{PET}. Such devices offer ever improving performance in terms of detection efficiency, time resolution, operation stability, etc.
\newline
Among others, the typical parameters that characterize a SiPM are its dark count rate and the amount of correlated noise that contributes to the detected counts. This latter consists of random counts initiated by additional photons that are produced in the development of the avalanche~\cite{Mirzoyan2009}. These additional counts can be either simultaneous with the primary avalanche (cross-talk) or they may be issued with a time delay due to the temporary trapping of such photons (after-pulses).
\newline
Important information about these quantities can be extracted from the noise charge spectrum $\frac{dN}{dQ}$, i.e. the charge distribution generated by the noise only, when the detector is not exposed to external light stimuli. A typical noise charge spectrum (from simulation) is shown in figure~\ref{fig: SPE}, where $Q$ indicates the measured charge, $Q_0$ is the charge produced by a single avalanche, $N_0$ is the number of measured signals.
A procedure that is commonly used to estimate the dark count rate and the contribution from the correlated noise is to consider the two quantities
\begin{equation}
  I_{0.5} = \int_{Q=0.5Q_0}^{\infty} \frac{dN}{dQ} dQ
\end{equation}
and
\begin{equation}
  I_{1.5} = \int_{Q=1.5Q_0}^{\infty} \frac{dN}{dQ} dQ,
\end{equation}
from which one tipically estimates the dark count rate as
\begin{equation}\label{DCR conv}
  DCR=\frac{I_{0.5}}{T},
\end{equation}
where $T$ is the total measurement time, while the amount of correlated noise would be given by the ratio
\begin{equation}\label{XT conv}
  x=\frac{I_{1.5}}{I_{0.5}}.
\end{equation}
Even though the method may give a good estimate, in general $DCR$ and $x$ as defined by formulae~\ref{DCR conv} and~\ref{XT conv} are only approximations of the true values, as it will be clear later. This may pose problems when, for example, one wants to use them as input parameters of a simulation. In fact, using equations~\ref{DCR conv} and~\ref{XT conv} assumes that the contribution of dark counts to the secondary peaks of the spectrum is negligible, whereas a consistent treatment of the problem requires considering dark counts and correlated noise counts as tightly entangled in producing the shape of the noise charge spectrum.
\newline
In this sense, the other common approach of fitting the noise charge spectrum with an analytical model (as, for example, in~\cite{FACT,Vinogradov2009}) is more correct. However, fitting may not always be the optimal solution. In particular, when the model presents many parameters, not only the convergence of the fit can become challenging, but also the uncertainties on the extracted values of the parameters (and so also on dark count and correlated noise rates) can be large due to correlations. Furthermore, some large systems may be built from a considerable number of SiPMs which may need to be characterized on a regular basis. For such systems, controlling the fit convergence could be challenging.
\newline
The aim of the present work is to formulate the problem of noise count generation in a SiPM in terms of a set of equations relying on a specific (and conventionally validated) model. The solution of the equations allows one to compute the dark count rate and the correlated noise contribution from quantities measured from the noise charge spectrum. Such a method avoids fitting and goes back to an approach based on analytical formulae where - however - the approximate expressions~\ref{DCR conv} and~\ref{XT conv} are replaced by exact solutions\footnote{relatively to the underlying model.}.
\newline
The paper is organized as follows. After introducing a general formulation (section~\ref{sec: general}), the statistical model is fixed to a branching Poisson process (section~\ref{sec: branching}) and the equations are solved numerically (section~\ref{sec: solution}). A discussion on the effect of statistical fluctuations follows, which is also useful to study the intrinsic measurement precision that is achievable as a function of the sample size (section~\ref{sec: fluct}). Finally, possible extensions to other statistical models are examined (section~\ref{sec: extensions}).


\section{General formulation of the problem}~\label{sec: general}
The statistics of noisy counts generation in a SiPM can be treated in terms of discrete probability distributions that depend on a few parameters. In the most general case, let's consider a set of $N_{DC}$ parameters $\vec{\lambda}=\lambda_1, \lambda_2, \ldots \lambda_{N_{DC}}$ that define a distribution $P_n(\vec{\lambda})$ describing the probability of generating $n$ dark counts. Another set of $N_{CN}$ parameters $\vec{\xi}=\xi_i, \xi_2, \ldots \xi_{N_{CN}}$ can be defined to describe the statistics of correlated noise via another distribution $\Pi_m(\vec{\xi})$, representing the probability of generating $m$ correlated noise counts. The physical interpretation of $\vec{\lambda}$ and $\vec{\xi}$ depends on the statistical model employed. The specific model that is developed in the following paragraphs depends on just two parameters: $\lambda$, from which the dark count rate can be derived, and $\xi$, which is related to the average number of correlated noise counts that are triggered by a primary event.
\newline
Be $N_0$ the total number of recorded \emph{signals} (not to be confused with the number of counts). $N_0$ thus represents the area of the full charge spectrum. And be $N_k$ the number of signals arising from $k$ \emph{simultaneous} counts. $N_k$ thus represents the area of the $k$-th peak of the charge spectrum. Counts are considered to be simultaneous when they occurr within a time window $\tau$, inside which the generated charges are summed up. Such time window is usually related to the pulse width of the analog signals produced by the SiPM. The proposed model is valid when one assumes that the characteristic time of after-pulses is small with respect to the time window $\tau$. After-pulses are thus treated in the same manner as cross-talk, and the term ``correlated noise'' is used to describe both processes without distinction.

\subsection{Noise spectrum equations}
The noise spectrum equations can be written down by considering all the possible cases for which events with $k$ counts can arise, with $k=1,2,\ldots$. Single count events are detected only when one dark count occurs and no correlated noise adds up. Two-count signals can be detected when either of the following two conditions are met: 1) two dark counts occur and no correlated noise; 2) one dark count occurs and one - and only one - additional count is generated as correlated noise. Following this reasoning, one may write a system of recursive equations
\begin{equation}\label{system 1}
  \begin{cases}
    N_1 = N_0 P_0(\vec{\lambda}) \Pi_0(\vec{\xi}) \\
    N_2 = N_0 \left[ P_1(\vec{\lambda}) \Pi_0(\vec{\xi}) + P_0(\vec{\lambda}) \Pi_1(\vec{\xi}) \right] \\
    \qquad \vdots \\
    N_k = N_0 \sum_{i=0}^{k-1} P_{k-1-i}(\vec{\lambda}) \Pi_i(\vec{\xi}) \\
    \qquad \vdots \\
  \end{cases}.
\end{equation}
Each equation defines the expected number of events falling in the $k$-th peak of the spectrum. In this notation, $P_n(\vec{\lambda})$ represents the pile-up probability, i.e. the probability that, given a dark count, $n$ additional counts from dark noise (and not from correlated noise) are detected simultaneously. At the same time, the correlated noise probabiliy $\Pi_m(\vec{\xi})$ represents the probability of having $m$ counts arising from correlated noise.
\newline
The system of equations~\ref{system 1} relates the unknowns $\vec{\lambda}$ and $\vec{\xi}$ to quantities that can be measured from the noise charge spectrum: its total area $N_0$ and the areas of its individual peaks $N_1,N_2,\ldots,N_k,\ldots$. Ideally, the possibility to invert such a system should lead to a solution of the form
\begin{equation}
  \begin{cases}
    \lambda_1 = f_1(N_0, N_1, \ldots) \\
    \qquad \vdots \\
    \lambda_{N_{DC}} = f_{N_{DC}}(N_0, N_1, \ldots) \\
    \xi_1 = f_1(N_0, N_1, \ldots) \\
    \qquad \vdots \\
    \xi_{N_{CN}} = f_{N_{CN}}(N_0, N_1, \ldots) \\
  \end{cases}
\end{equation}
so that the $\vec{\lambda}$ and $\vec{\xi}$ parameters can be directly calculated from the measured quantities.
\newline
The main idea is that, although in the general formulation one can write equations for any value of $k$, when dealing with specific models of dark counts and correlated noise, the system of equations is overdetermined and just a few equations, possibly the first (and simplest) ones, are needed. This concept is developed in the following section for the case of a branching Poisson model.


\section{Branching Poisson model}\label{sec: branching}
A generally accepted model to describe the statistical generation of dark signals in a SiPM is the one considering a branching Poisson process~\cite{Vinogradov2009,Vinogradov2012}. In such a process, the correlated noise arises from the chain generation of Poisson distributed counts.

\subsection{Poisson distribution of dark counts}
Due to the cell structure of a SiPM, the statistics of dark counts is intrinsically binomial. However, since the number of cells is usually high and the dark count rates are small enough that only a small fraction of the cells are fired simultaneously within a gate window $\tau$, the Poisson limit theorem may be applied and the firing of cells can be considered on a very good approximation as Poisson distributed. One may thus specify
\begin{equation}\label{P_k}
  P_k(\lambda) = \frac{\lambda^k e^{-\lambda}}{k!},
\end{equation}
with
\begin{equation}\label{lambda def}
  \lambda = 2 R \tau,
\end{equation}
where $R$ is the dark count rate. The most general vector of parameters $\vec{\lambda}$ introduced earlier thus reduces to a single parameter, $\lambda$, and the generic equation in the system~\ref{system 1} takes the form
\begin{equation}\label{general}
        n_k = e^{-\lambda} \sum_{i=0}^{k-1} \frac{\lambda^{k-1-i}}{(k-1-i)!} \Pi_i(\xi),
\end{equation}
where, for convenience, one has defined the quantity
\begin{equation}
  n_k = \frac{N_k}{N_0},
\end{equation}
that is the fraction of recorded signals of charge index $k$, or the fractional area of the $k$-th peak of the charge spectrum.

\subsection{Borel model of correlated noise}
As shown in~\cite{Vinogradov2012,Vinogradov2009}, a possible model to describe the generation of correlated noise is by a Borel distribution
\begin{equation}\label{Pi_k}
        \Pi_{k}(\xi) = B_{k+1}(\xi) = \frac{((k+1)\xi)^{k} e^{-(k+1)\xi}}{(k+1)!}.
\end{equation}
Such a model depends on a single parameter - $\xi$ - taking values in the interval $[0,1]$ and representing the average number of correlated counts that are generated at each step of the chain. The expected value
\begin{equation}
  \mu = \frac{1}{1-\xi}
\end{equation}
then gives the average number of correlated noise counts that are generated over the full chain.

\subsection{Equations for the Borel model}
Since the number of parameters has been reduced to two ($\lambda$ and $\xi$), one may try and consider just two equations to be solved simultaneously. By choosing the two equations obtained from~\ref{general} for $k=0$ and $k=1$, the system may be reduced to
\begin{equation}\label{system 2}
        \begin{cases}
                \lambda = n_{21} - \xi e^{-\xi} \\
                \xi \left( e^{-\xi} -1 \right) = n_{21} + \log n_1
        \end{cases},
\end{equation}
where
\begin{equation}
  n_{21} = \frac{N_2}{N_1},
\end{equation}
is the ratio between the areas of first two peaks in the charge spectrum.
\newline
Although the system~\ref{system 2} is not in diagonal form, its second equation only contains the unknown $\xi$. Once this equation is solved, the value of $\lambda$, and therefore the dark count rate $R$ (according to equation~\ref{lambda def}), is given directly from the first equation. This means that a SiPM obeying this statistical model can be fully characterized once $N_0$, $N_1$ and $N_2$ are measured from the charge spectrum and the system~\ref{system 2} is consequently solved (numerically or by approximation) for $\lambda$ and $\xi$. Further peaks of the spectrum (which, in general, are accounted for in a fitting model) give redundant information.


\section{Numerical and approximate solutions}\label{sec: solution}
The solution of the second equation in system~\ref{system 2} requires either numerical or approximation methods. The numerical solution can be found by finding the zero of the quantity
\begin{equation}
  L(\xi) = \xi \left( e^{-\xi} -1 \right) - n_{21} - \log n_1
\end{equation}
with respect to $\xi$ once $n_1$ and $n_{21}$ are calculated from $N_0$, $N_1$ and $N_2$. On the other hand, by expanding the exponential and retaining the first term in $\xi$, one obtains the approximate solution
\begin{equation}
  \xi \sim \sqrt{ - n_{21} - \log n_1 }.
\end{equation}
\begin{figure}
  \centering
  \includegraphics[width=0.5\textwidth]{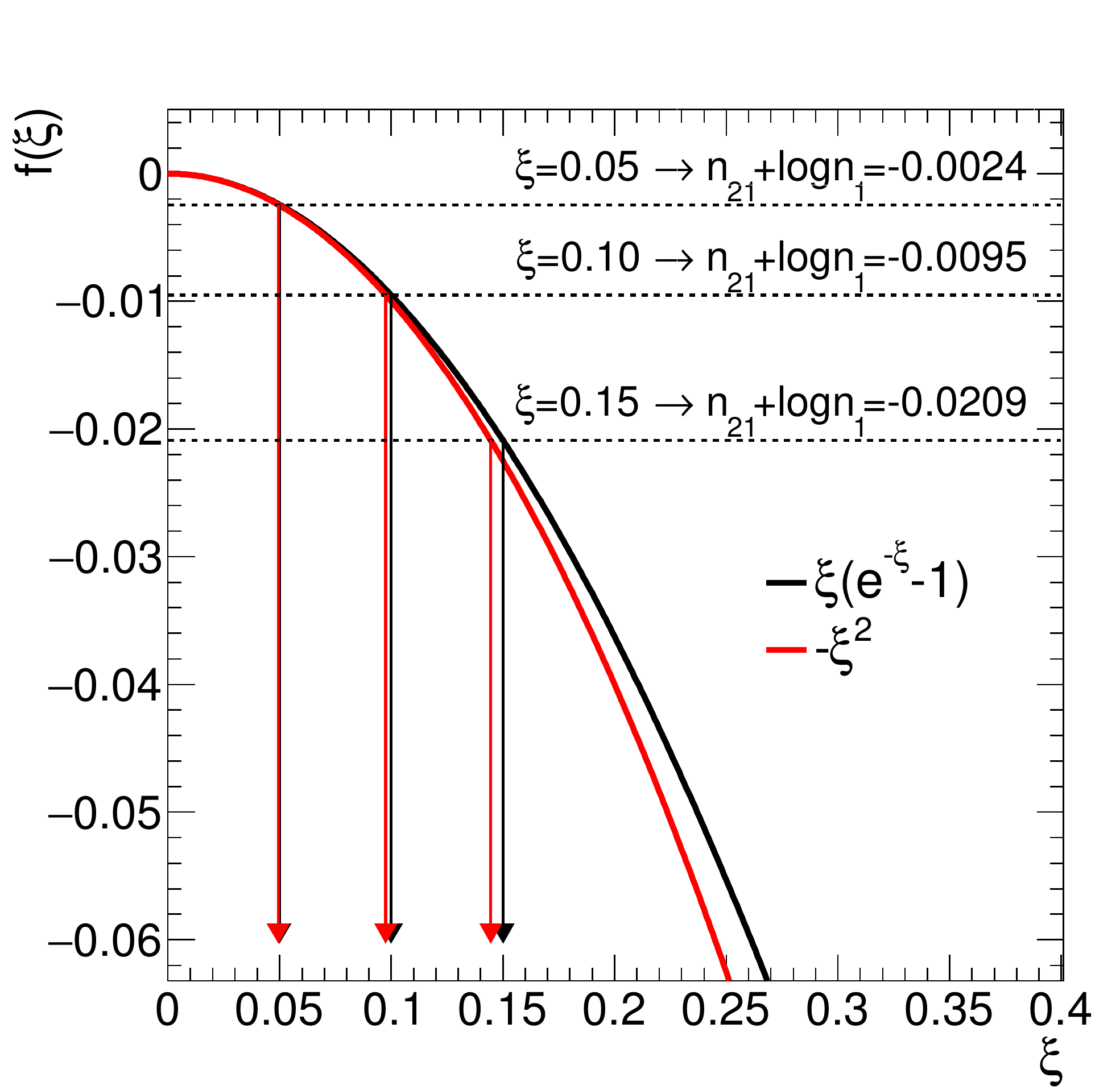}
  \caption{A few solutions (exact in black and approximate in red) of the noise spectrum equation for the Borel model.}
  \label{fig: solution}
\end{figure}
Figure~\ref{fig: solution} shows an example (for $\lambda=0.03$, correspoding to $R=1$~MHz if $\tau=15$~ns) of a few solutions found with either methods. The numerical method is equivalent to finding the intersections between
\begin{equation}\label{f(xi)}
  f(\xi) = \xi \left( e^{-\xi} -1 \right),
\end{equation}
represented by the plain black curve, and the constants
\begin{equation}
  c =  n_{21} + \log n_1,
\end{equation}
represented by the dotted horizontal lines calculated for different values of $\xi$. At the same time, the leading order approximate solution is equivalent to finding the intersection of the same constants with the expansion
\begin{equation}
  f(\xi) \sim -\xi^2.
\end{equation}
This latter corresponds to the red plain curve in the plot. The black and red vertical arrows point towards the values of $\xi$ that solve the equation in either cases: the numerical solutions return the exact values, the approximate solution return values that are relatively accurate only when $\xi$ is close to zero, e.g. $\xi\sim$~0.05.


\section{Influence of statistical fluctuations}\label{sec: fluct}
The statistical fluctuations on $N_0$, $N_1$ and $N_2$~\footnote{for a generic model, on each $N_k$, $k=0,1,2,\ldots$} are propagated to the $n_1$ and $n_{21}$ parameters, and therefore, when solving the system of equations~\ref{system 2}, they produce a distribution for the solutions $\xi$ and $\lambda$ rather than single values as in figure~\ref{fig: solution}. A systematic treatment of this effect is useful to establish the level of uncertainty that can be achieved in the measurement of dark count rate and correlated noise as a function of the size of the measured sample. Or, conversely, one may thus calculate how much statistics is needed if such quantities have to be measured with a given precision.
\subsection{Numerical evaluation}
The effect of the statistical fluctuations has been evaluated numerically. A typical measurement has been considered where one collects a fixed number $N_0$ of signals and produces a charge distribution. The first two peaks of the distribution are populated by a number of events $N_1$ and $N_2$, given by the first two equations in the system~\ref{system 1} after choosing $P_k(\lambda)$ and $\Pi_{k}(\xi)$ according to definitions~\ref{P_k} and~\ref{Pi_k}, respectively. In this framework, $N_1$ and $N_2$ are binomially distributed. Since one considers a large number of events, Gaussian distributions can be employed centered at $N_k$ and whose width is given by the square root of the binomial variance $\sqrt{N_k\left(1-\frac{N_k}{N_0}\right)}$, $k=1,2$. These distributions have been propagated in the calculation of the $n_{21} + \log n_1$ term, that is the right hand side of the second equation in the system~\ref{system 2}. The inverse image via $f(\xi)$ (equation~\ref{f(xi)}) of the distribution thus produced yields the distribution of $\xi$. The distribution of the dark count rate $R$ can be calculated with the same principle by applying the first equation of the system~\ref{system 2}, followed by equation~\ref{lambda def}, to the $\xi$  distribution.
\begin{figure}
  \centering
  \includegraphics[width=0.3\textwidth]{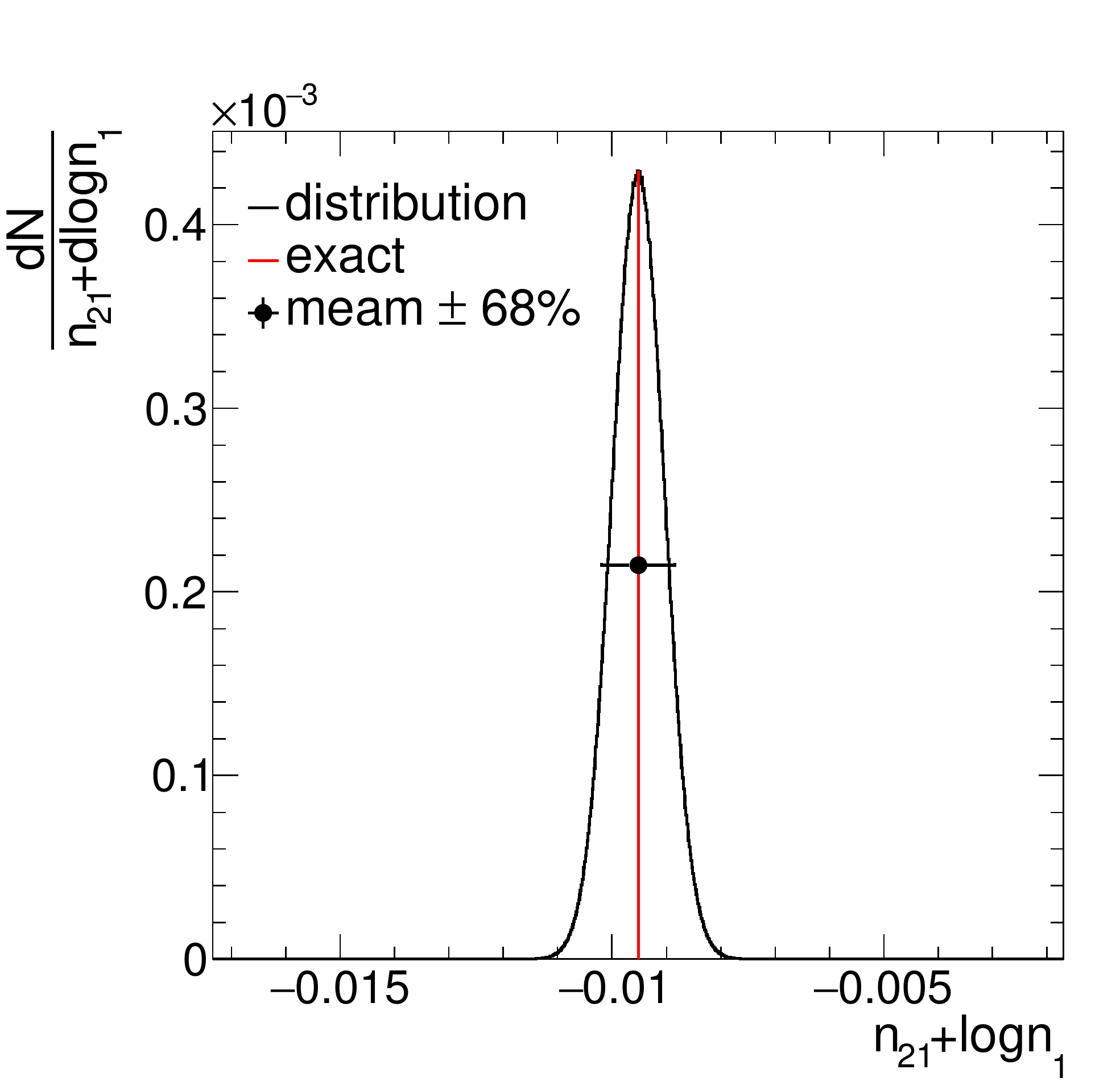}
  \includegraphics[width=0.3\textwidth]{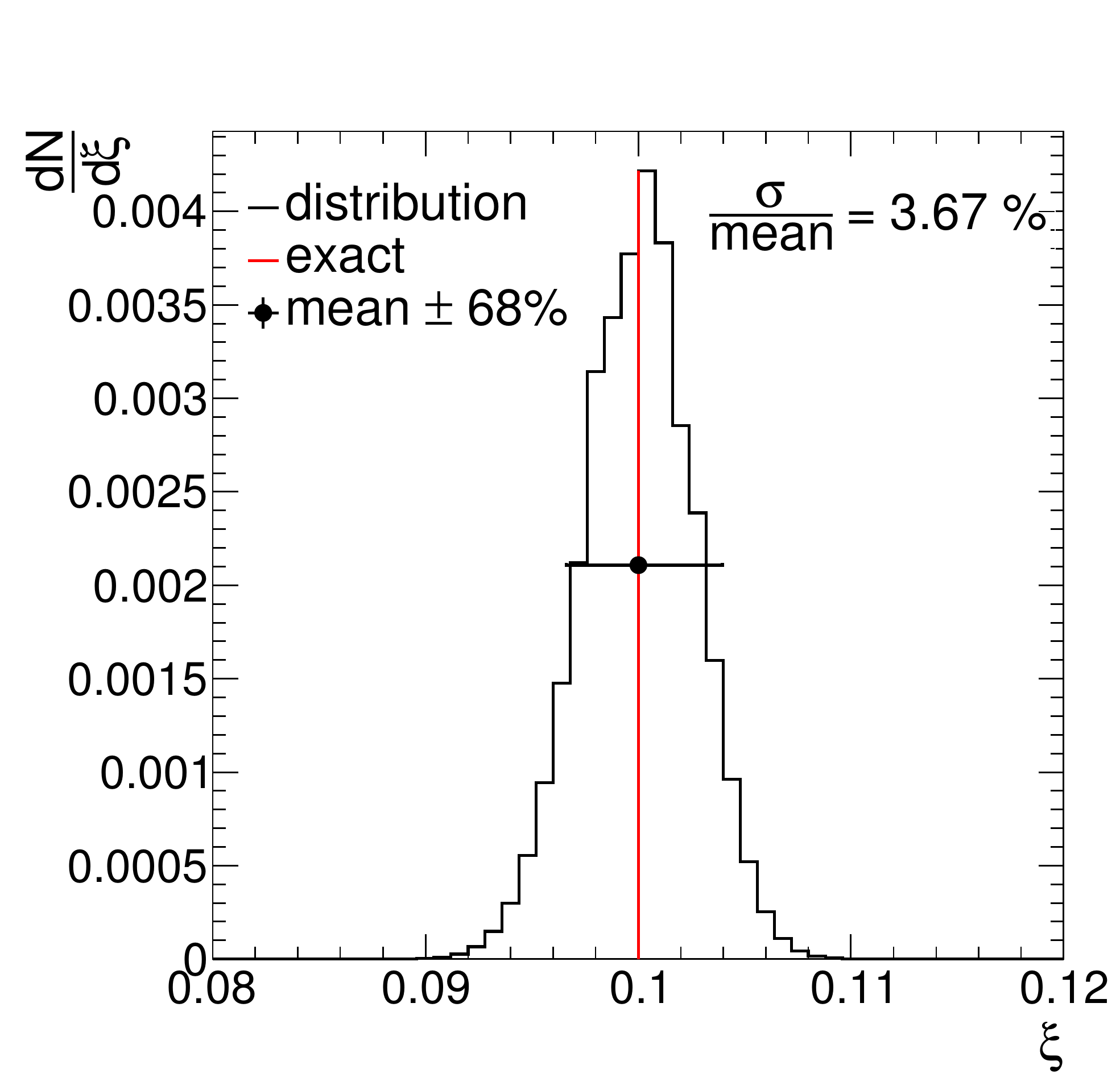}
  \includegraphics[width=0.3\textwidth]{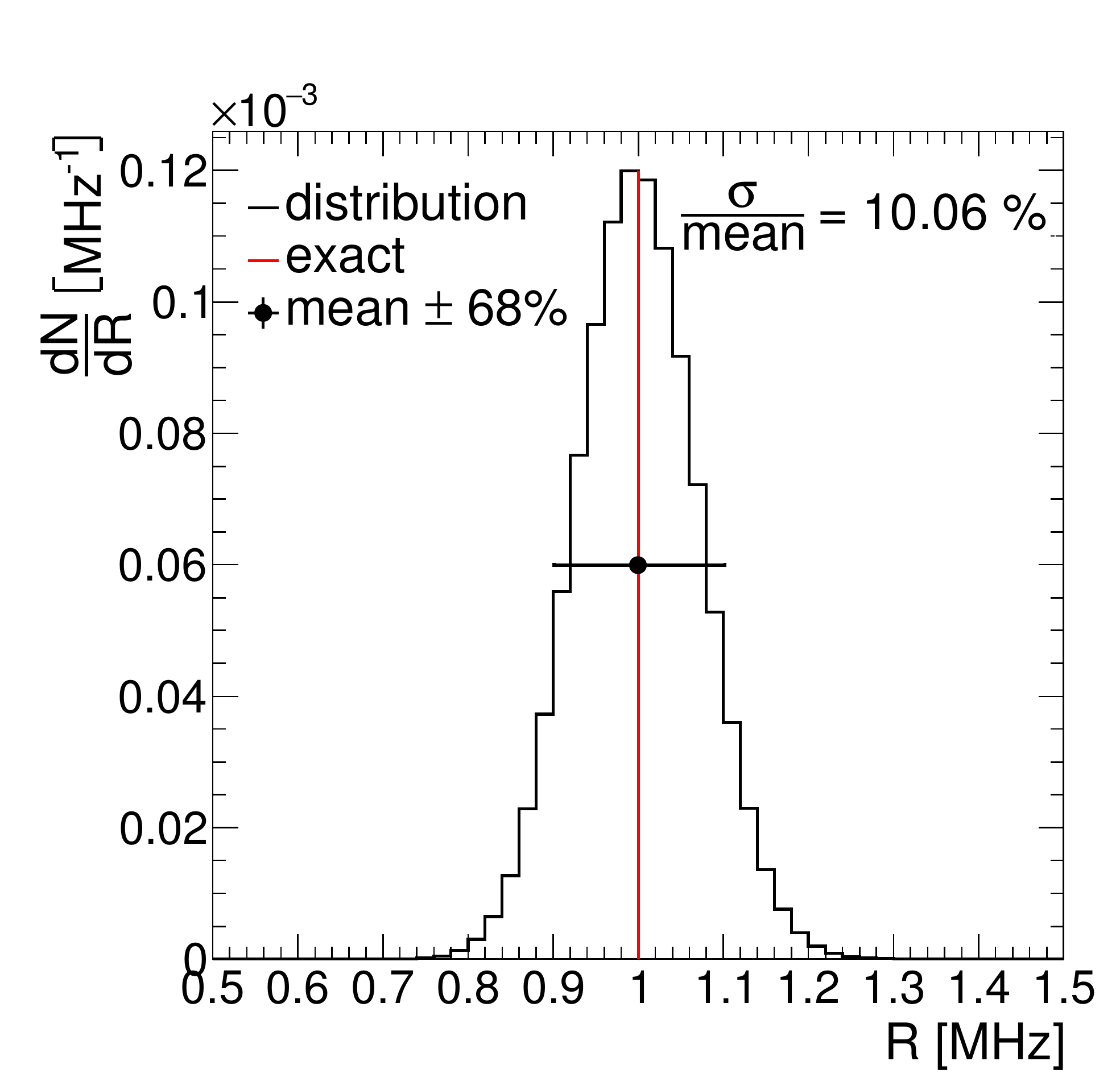}
  \caption{Example of a numerical calculation of (from left to right) $n_{21} + \log n_1$, $\xi$ and dark count rate distributions (solid black), compared to the exact values (solid red). The single markers show the mean values and their associated uncertainty, calculated as the 68\% area of the distribution.}
  \label{fig: statistics}
\end{figure}
An example of such a numerical calculation is shown in figure~\ref{fig: statistics}, produced by taking $N_0=10^5$, $\xi=0.1$, $R=1$~MHz and $\tau=15$~ns (i.e. $\lambda=$~0.03).

\subsection{Effects on undersized samples}
\begin{figure}
  \centering
  \includegraphics[width=0.4\textwidth]{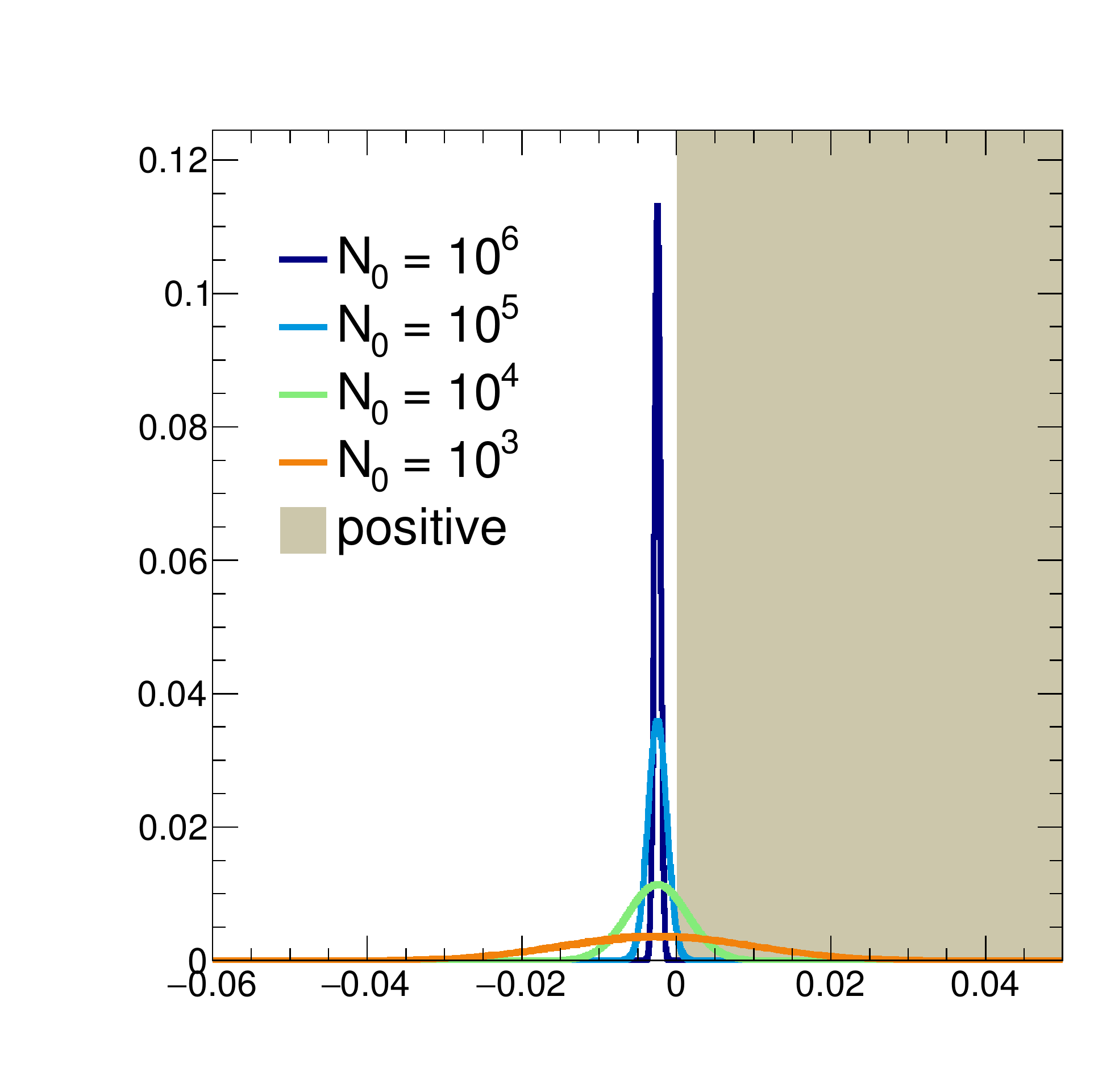}
  \includegraphics[width=0.4\textwidth]{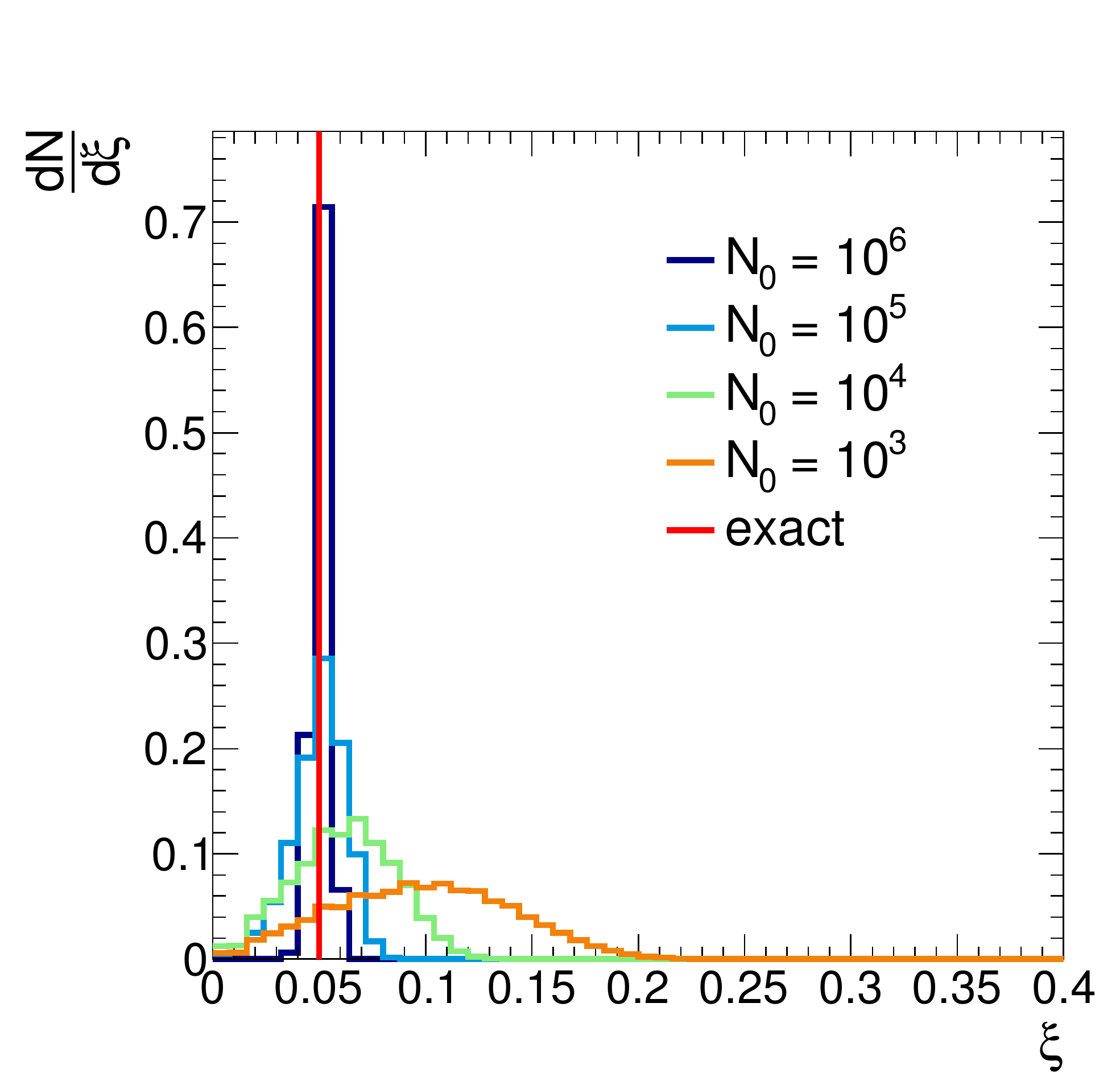}
  \caption{Numerical calculation of the effect of the statistical fluctuations on samples of different sizes (different colors in the plots). Left: the distribution of $n_{21} + \log n_1$. Events falling in the positive semi-axis (shaded region) cannot be used to solve the noise spectrum equations. Right: distribution of $\xi$, where the red vertical line indicates the true value of $\xi$.}
  \label{fig: low XT}
\end{figure}
\begin{figure}[h!]
  \centering
  \includegraphics[width=0.3\textwidth]{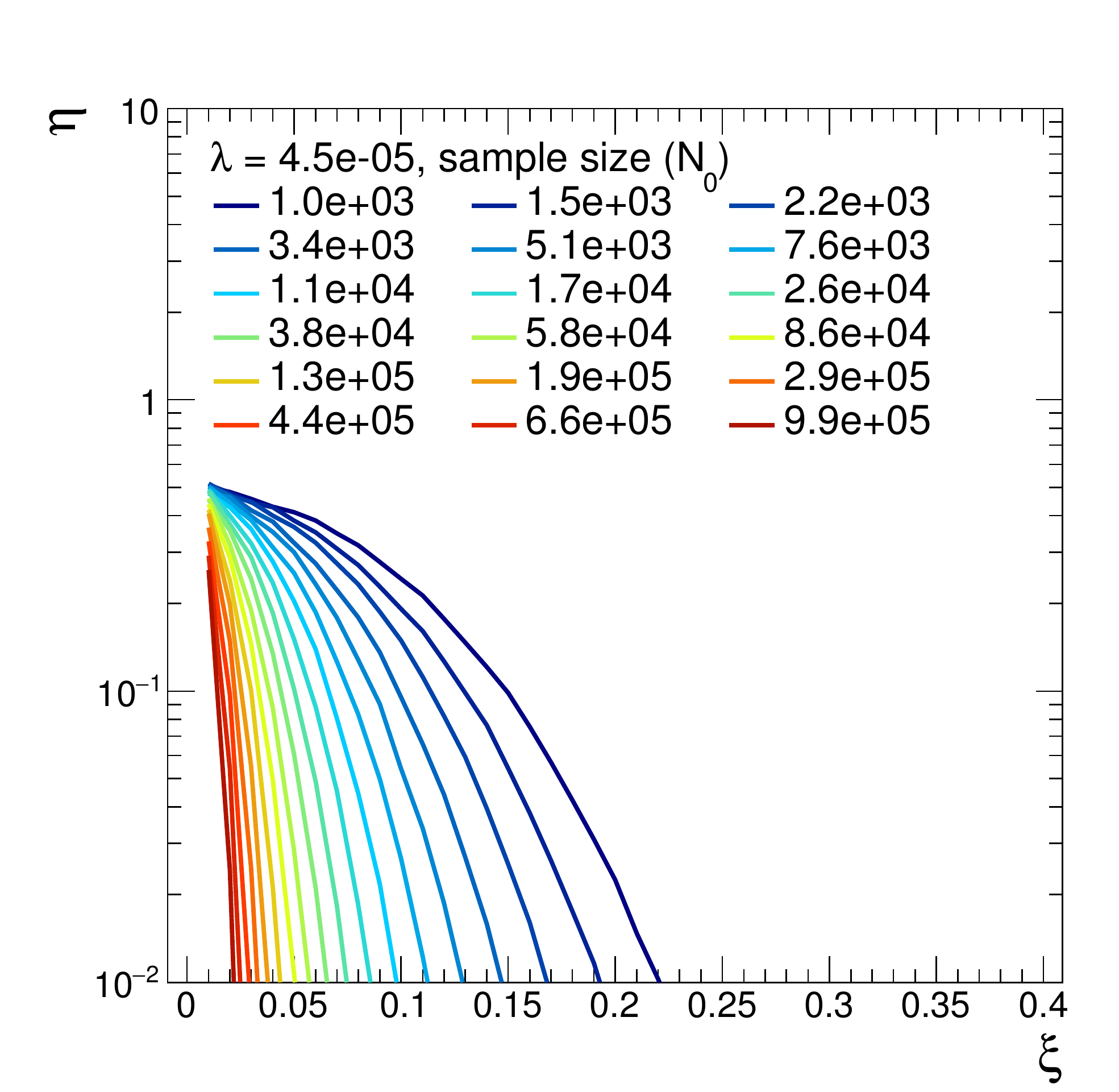}
  \includegraphics[width=0.3\textwidth]{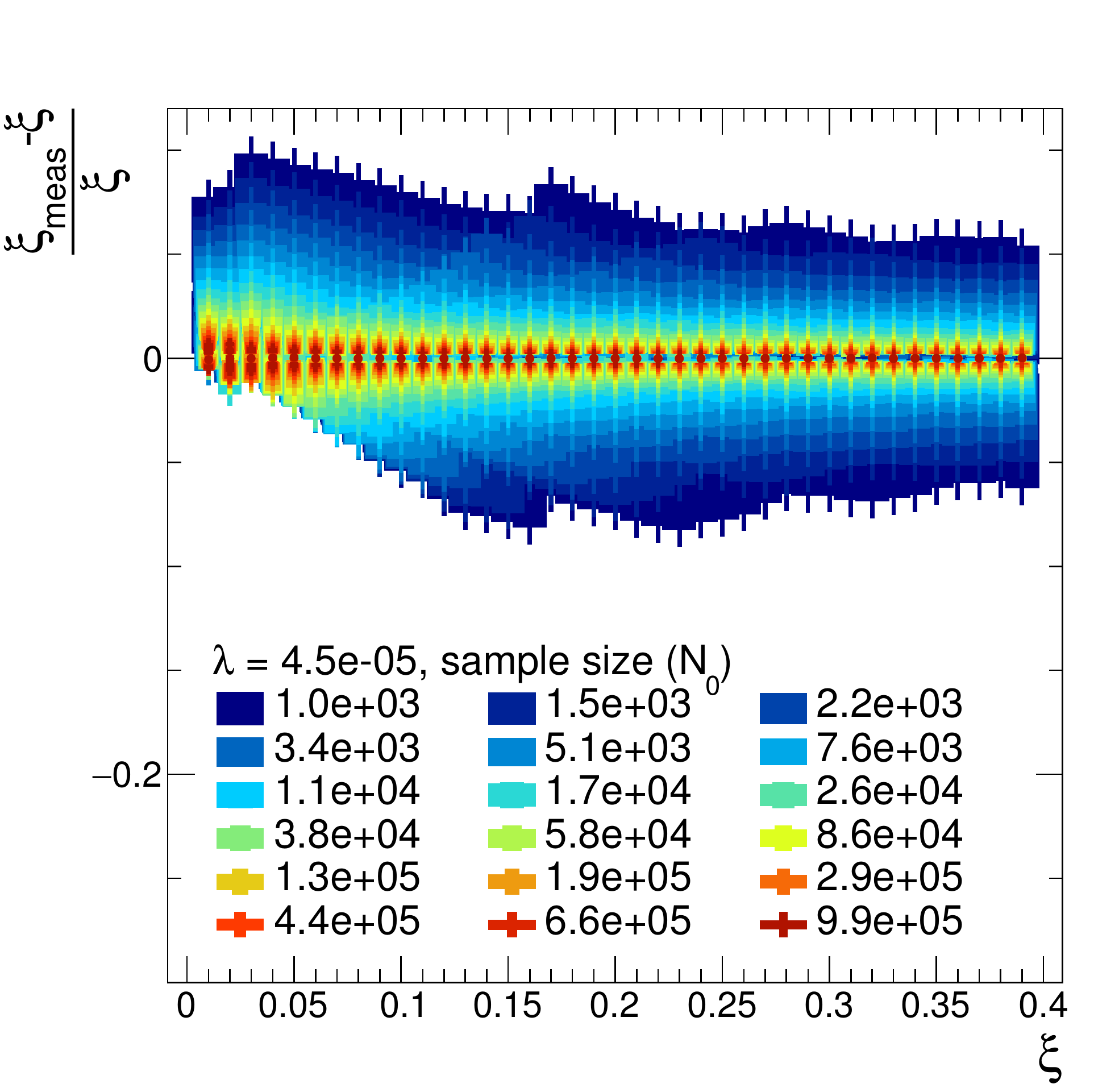}
  \includegraphics[width=0.3\textwidth]{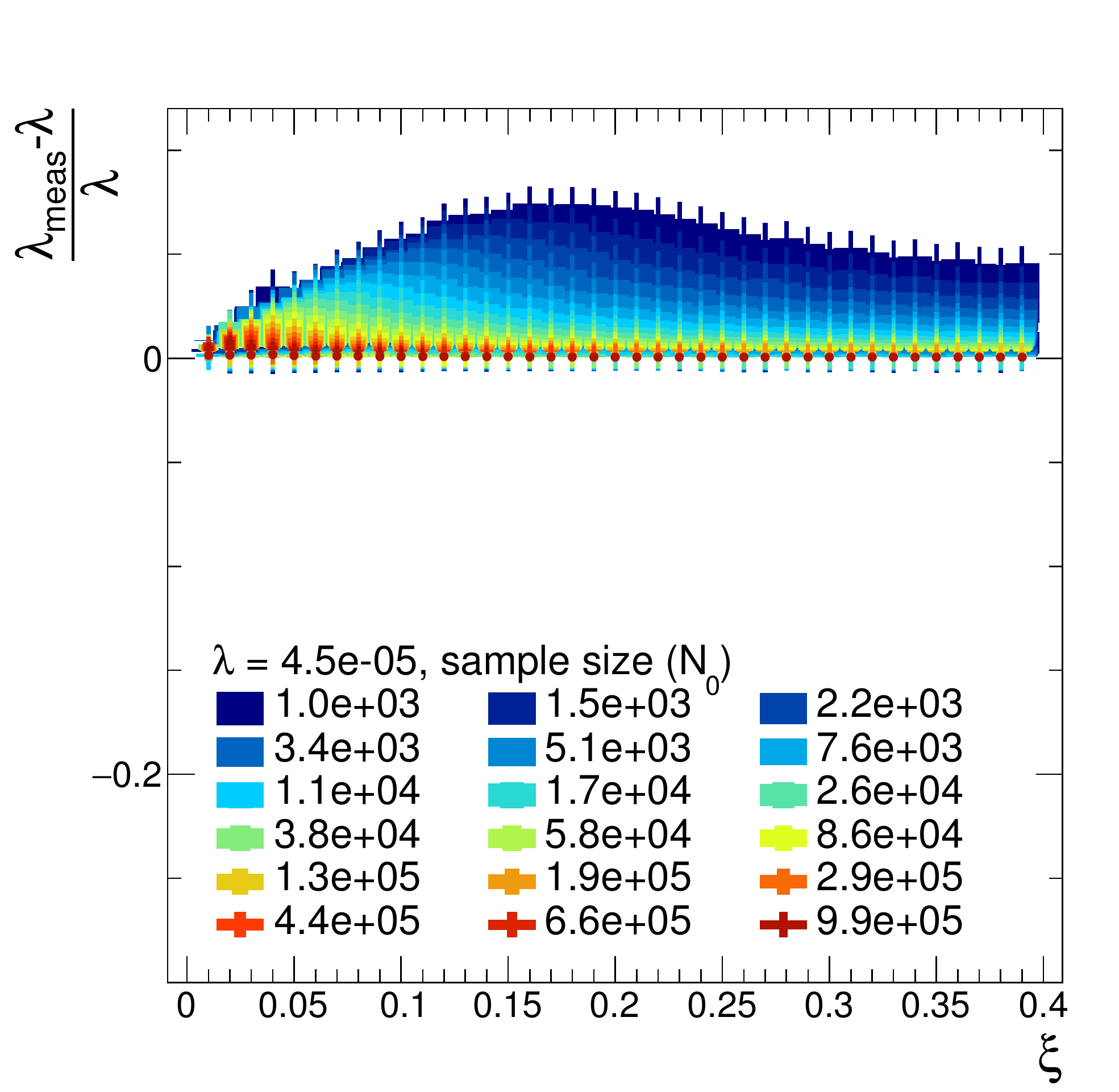}
  \includegraphics[width=0.3\textwidth]{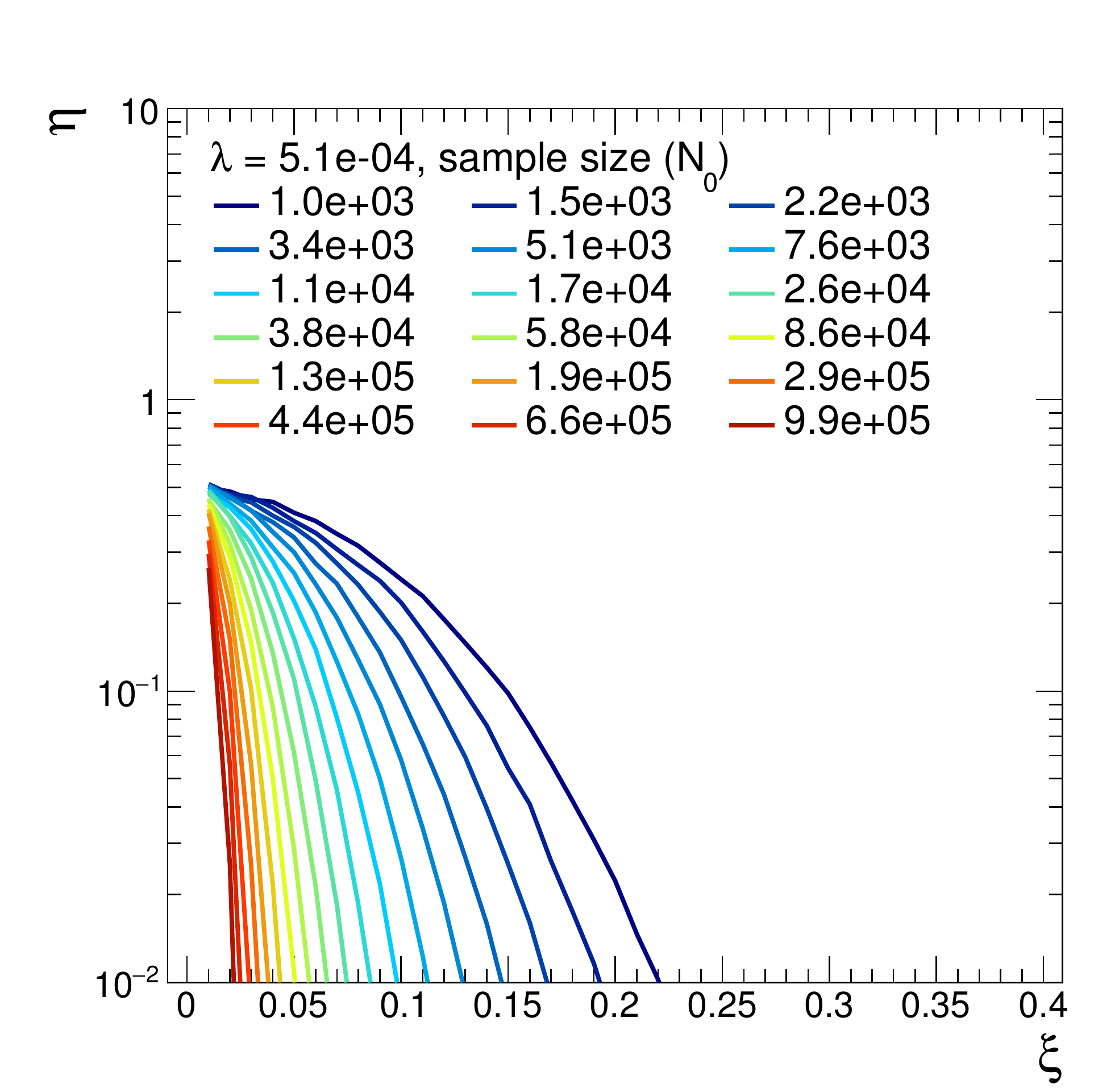}
  \includegraphics[width=0.3\textwidth]{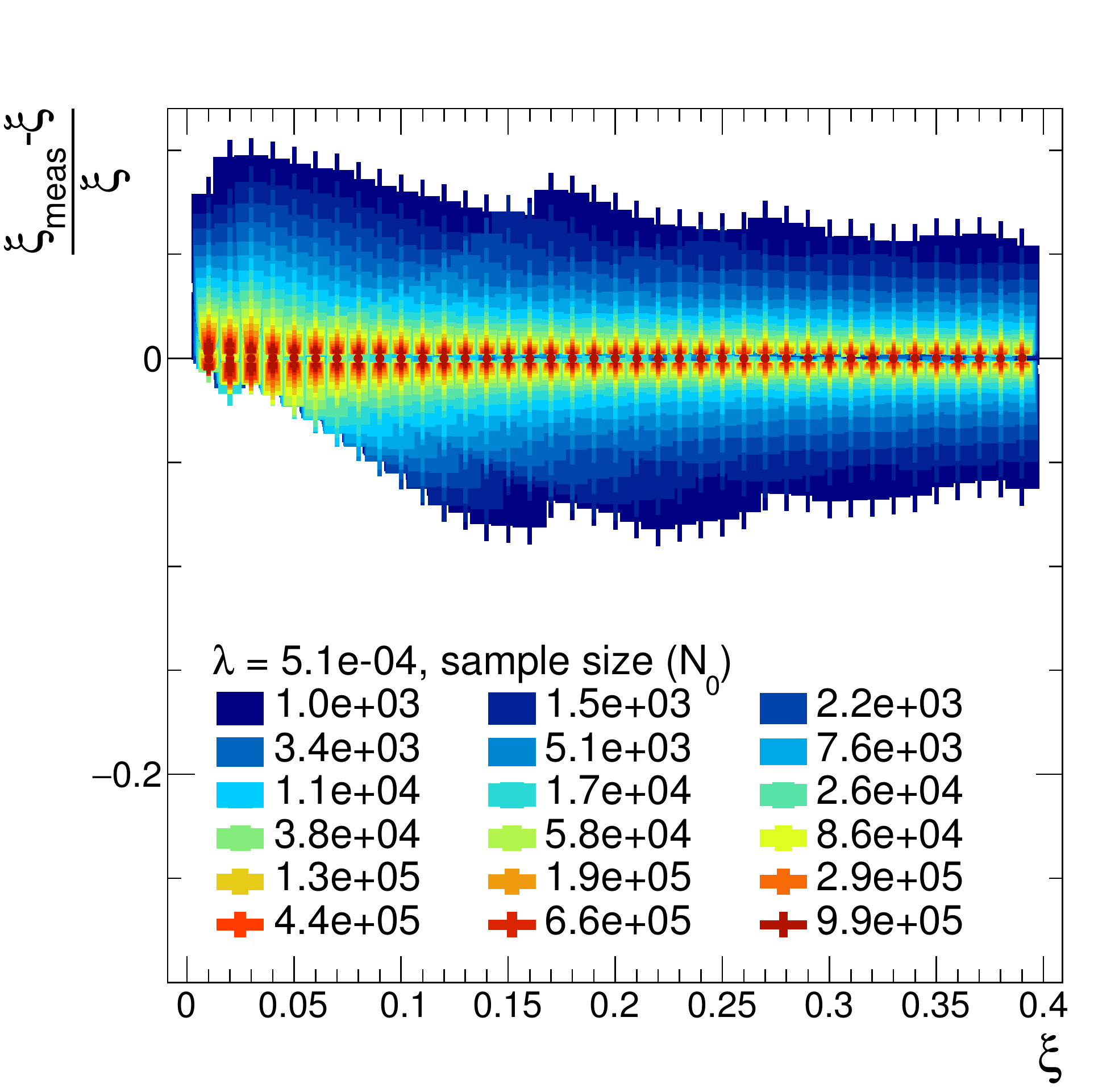}
  \includegraphics[width=0.3\textwidth]{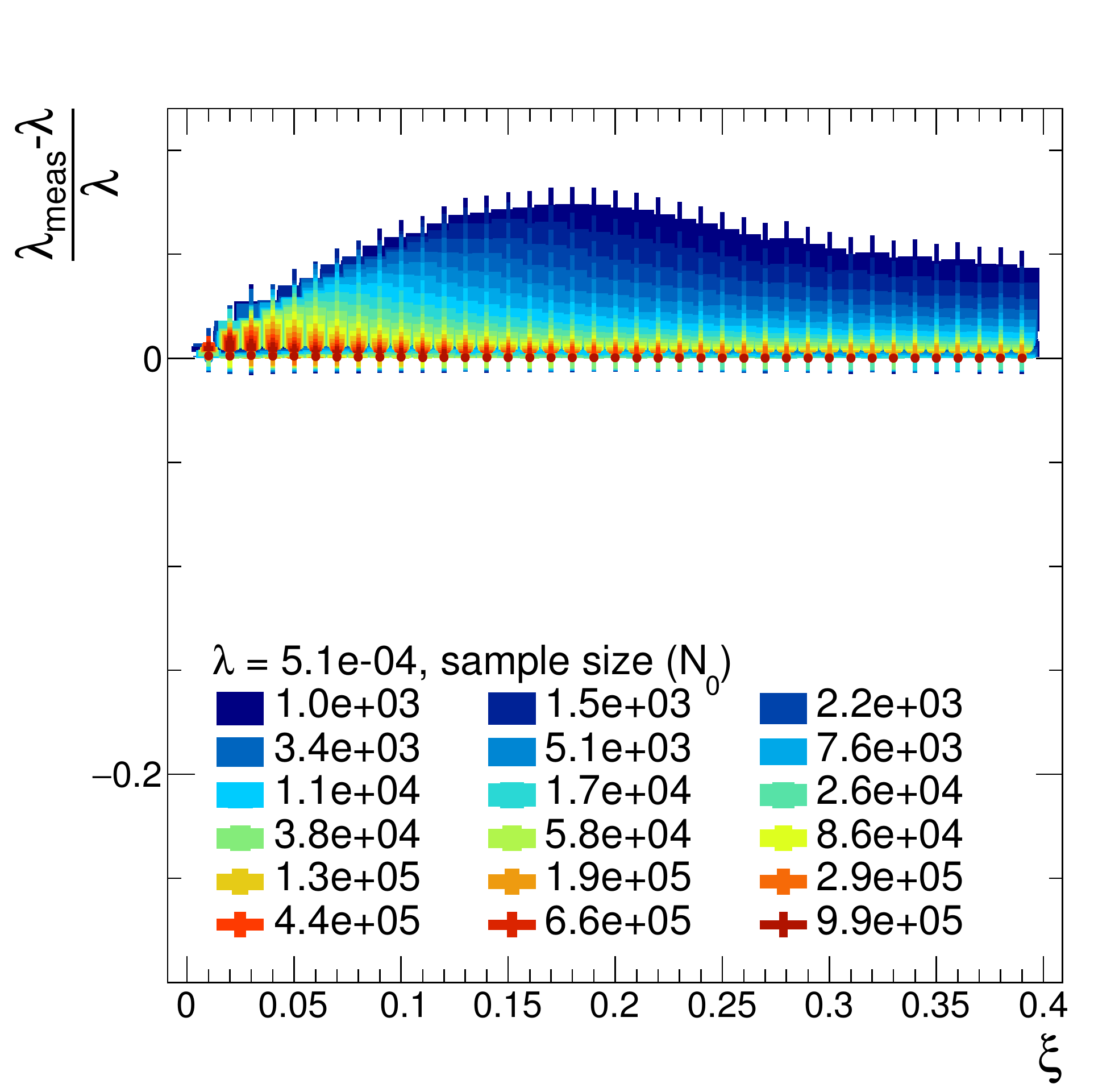}
  \includegraphics[width=0.3\textwidth]{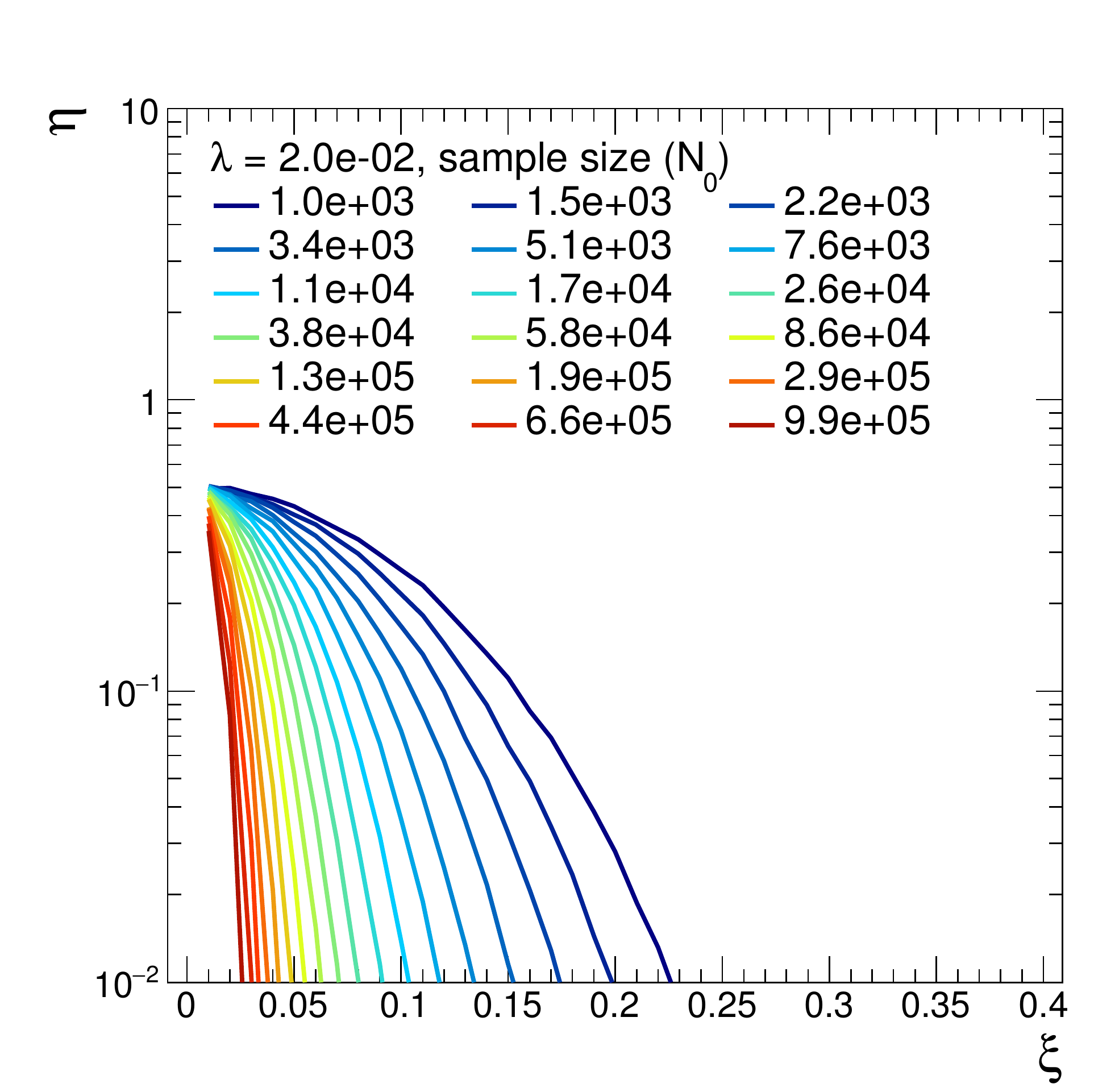}
  \includegraphics[width=0.3\textwidth]{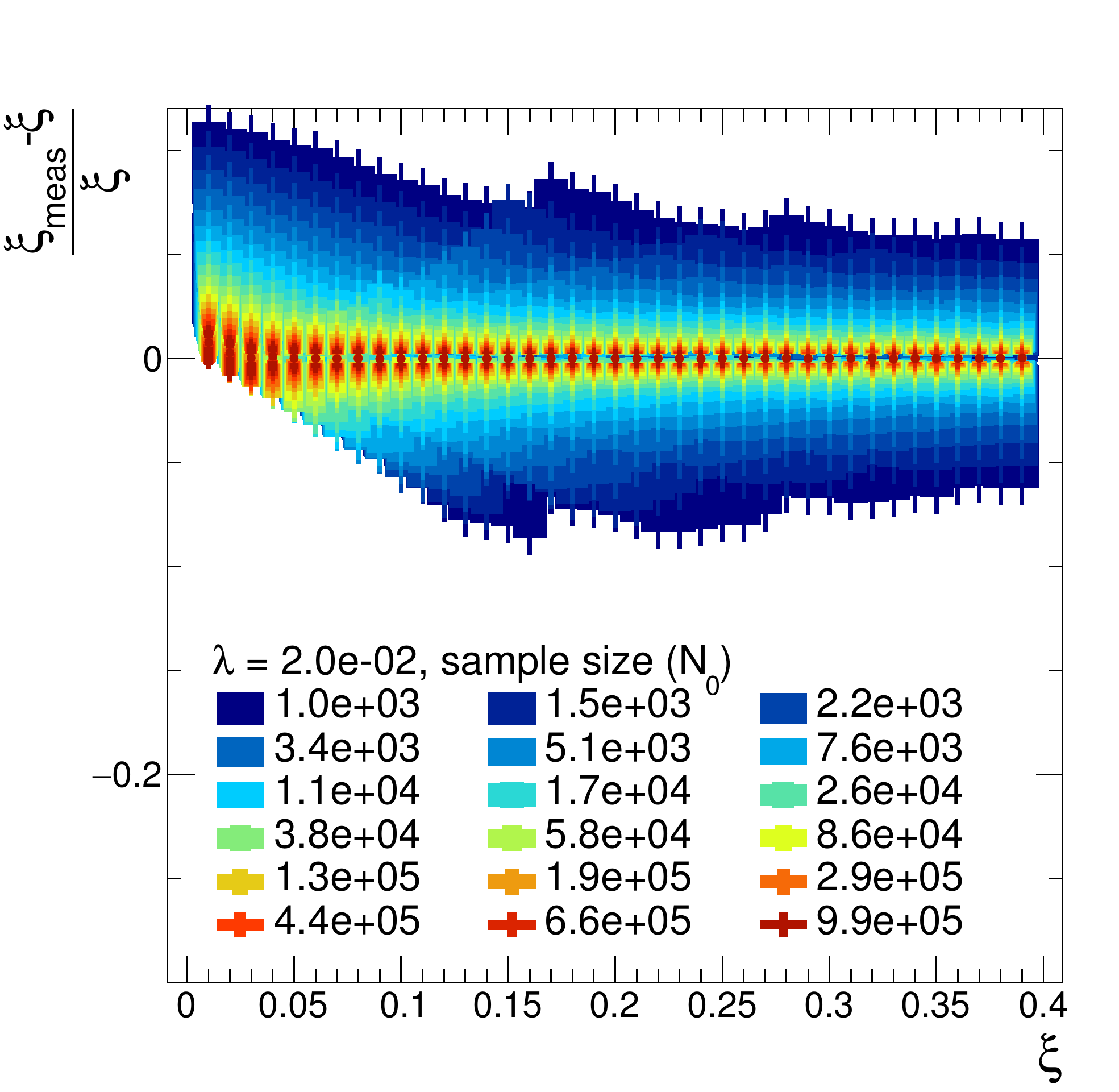}
  \includegraphics[width=0.3\textwidth]{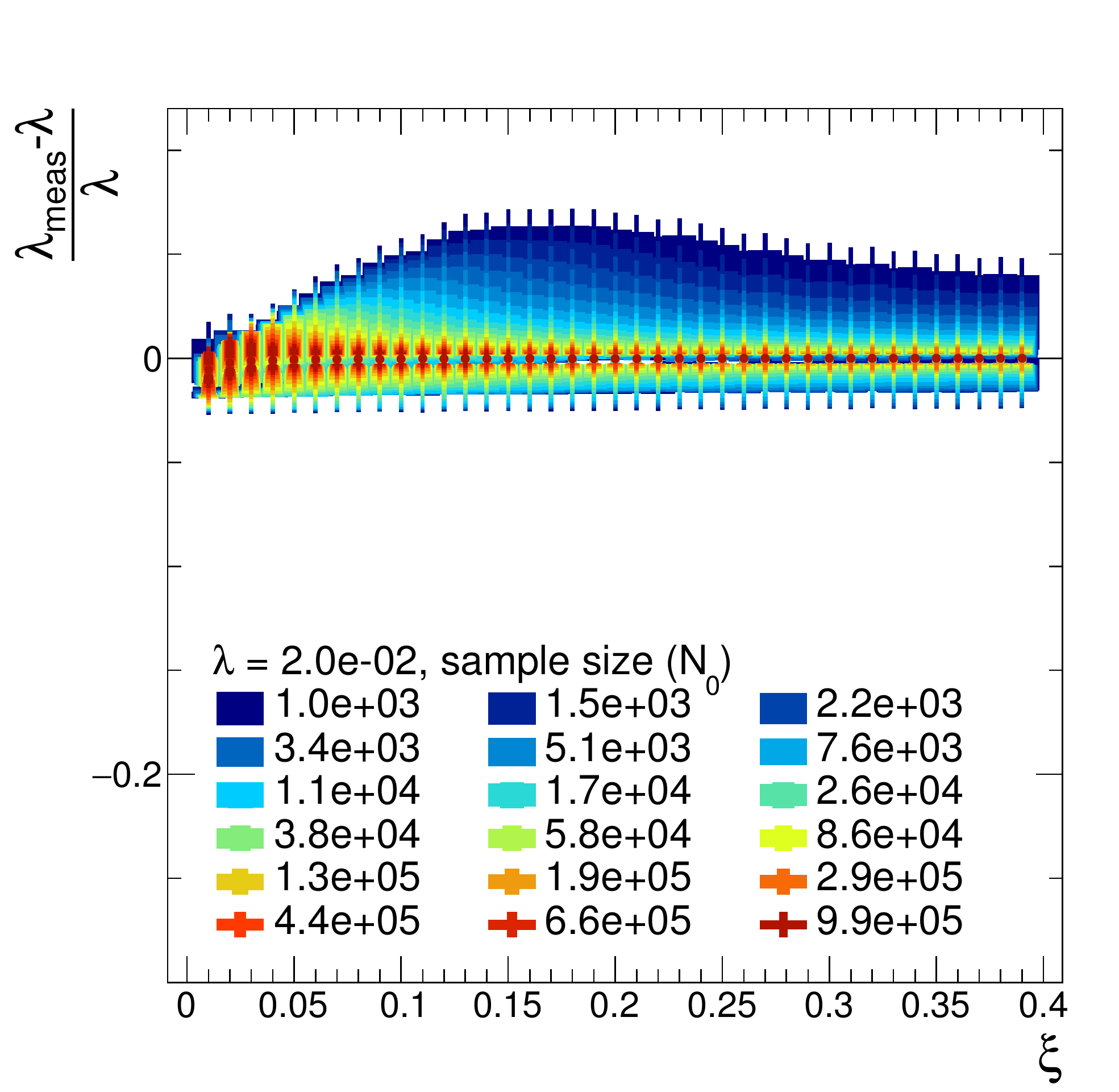}
  \includegraphics[width=0.3\textwidth]{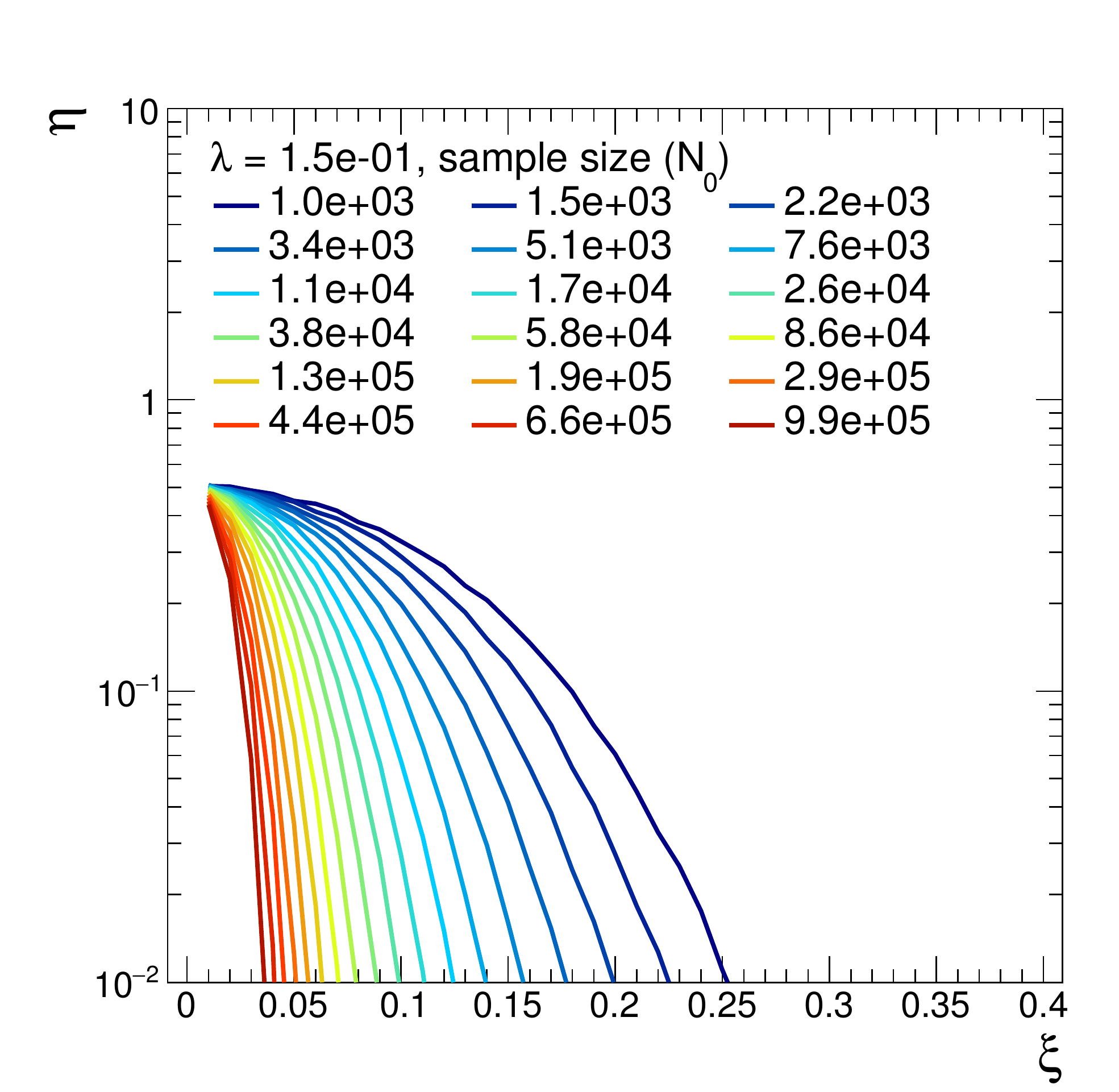}
  \includegraphics[width=0.3\textwidth]{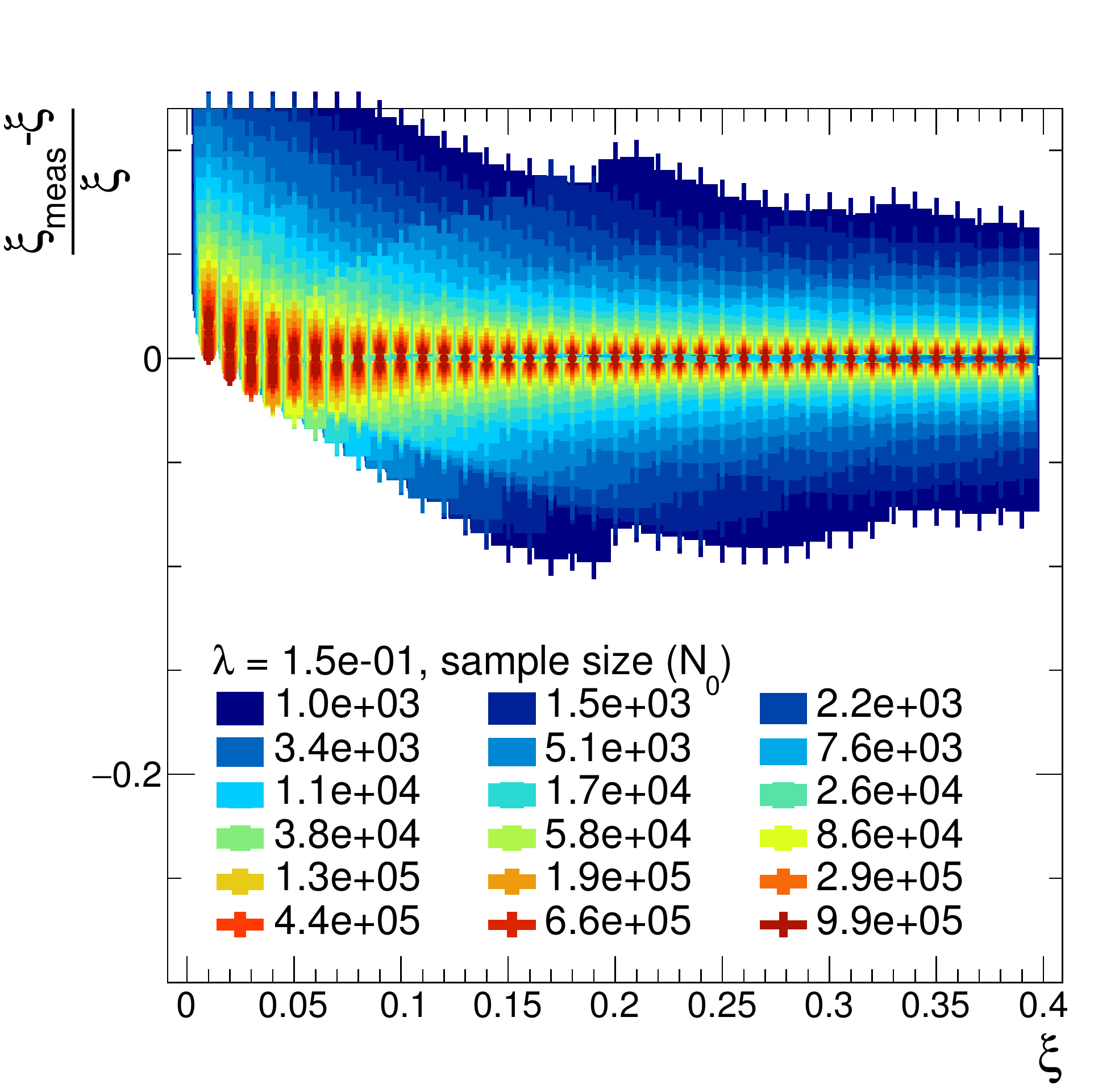}
  \includegraphics[width=0.3\textwidth]{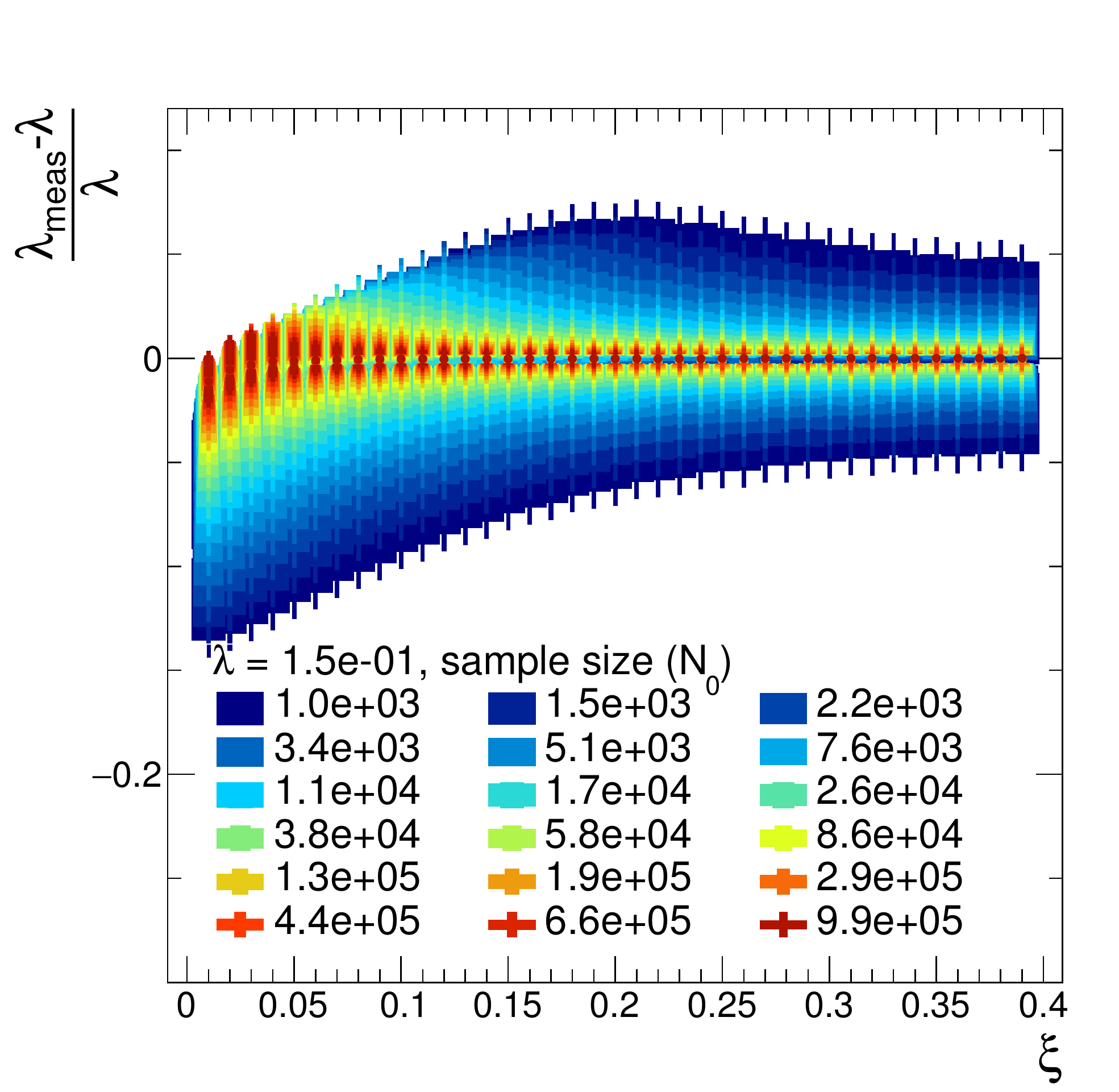}  
  \caption{In each row, as a function of $\xi$: (left) $\eta$, i.e. the fraction of the $n_{21} + \log n_1$ distribution falling on the positive axis, (center) relative deviation of the measured $\xi$ from its true value, (right) relative deviation of the measured $\lambda$ from its true value. The measured values of $\xi$ and $\lambda$ are calculated as the most probable values of their distributions. Error bands are calculated as 68\% of the areas of such distributions. From top to bottom: $\lambda = 4.5\times10^{-5}$, $\lambda = 5.1\times10^{-4}$, $\lambda = 2.0\times10^{-2}$ and $\lambda = 1.5\times10^{-1}$. For $\tau=15$~ns, such values correspond to $R=1.5$~kHz, $R=17$~kHz, $R=667$~kHz and $R=5$~MHz. Structures in the error bands at low $N_0$'s are artifacts due to the discretization of the numerical calculation.}
  \label{fig: eta}
\end{figure}
A necessary condition for the solution of~\ref{system 2} to exist is that $n_{21} + \log n_1$ must be negative. However, due to the statistical fluctuations on $N_1$ and $N_2$, in general the distribution of such a quantity has support over the full real axis. This implies that a finite fraction of the distribution is defined on the positive axis and cannot be used to solve the system. Such a fraction depends on the sample size $N_{0}$, which determines the width and shape of the $n_{21} + \log n_1$ distribution (see figure~\ref{fig: low XT}, left). As a consequence, if the sample is undersized, the $\xi$ distribution (and, therefore, also the $R$ distribution) can be severely distorted (figure~\ref{fig: low XT}, right), and the chance is lower of obtaining the correct values of $\xi$ and $R$ from a single measurement. The effect is stronger the smaller the correlated noise, due to the flattening of $f(\xi)$ close to $\xi=0$.
\newline
The example shown in figure~\ref{fig: low XT} is useful to visualize such an effect. A calculation has been carried out for $\xi=0.05$, $R=1$~MHz and $\tau=15$~ns ($\lambda=$~0.03) at different sample sizes ($N_0=10^3,10^4,10^5,10^6$). The smaller $N_0$, the larger the fraction of the $n_{21} + \log n_1$ distribution falls on the positive axis (left plot), the larger the distortion of the $\xi$ distribution (right plot) and the larger the possibility that $n_{21} + \log n_1$ is non-negative (and thus $\xi$ and $R$ cannot be measured), or that the value of $\xi$ that is measured deviates significantly from the true value. This effect is particularly visible in the $\xi$ distribution for $N_0=10^3$, which is significantly de-centered from the true value.
\newline
Calculations of this type have been carried out systematically to determined the sample size required to measure correlated noise and dark count rate with a given precision. A few examples of the calculation results are shown in figure~\ref{fig: eta}. These show that, as expected, one can reach an arbitrary precision by choosing appropriately large samples. However, for a given sample size, measurements are less precise the lower $\lambda$ and/or $\xi$ are. In particular, the $\lambda$ distributions (and therefore the $R$ distributions) may be strongly asymmetric at low dark count rates.


\section{Possible extensions of the model}\label{sec: extensions}
While the assumption that dark counts are Poisson distributed is generally correct, and therefore equations~\ref{general} are generally valid, the Borel model employed to derive equations~\ref{system 2} may be appropriate only for some devices, while for others the generation of correlated noise may be better described by other stochastic processes, possibly depending on more parameters than simply $\lambda$ and $\xi$ (see, for example,~\cite{FACT-calibration} or~\cite{Gallego}). Following the reasoning presented here for the Borel case, once the model is established, a specific set of (possibly few) equations may be derived from the most general system~\ref{system 1} or~\ref{general}. The characteristic parameters of the SiPM can be calculated from the areas of a few peaks and the total area of the noise charge spectrum via a diagonalization of the system.


\section{Conclusion}
In the proposed formulation, the stochastic generation of dark counts and correlated noise counts is treated in terms of a set of recursive equations. For a specific statistical model in which dark count is Poisson distributed and correlated noise is described by a branching Poisson process via a Borel distribution, the full system reduces to just two equations in two unknowns, $\lambda$ and $\xi$, which are related  to the dark  count rate and the average correlated noise counts. The system is solved numerically to show how the dark count rate and the average number of correlated noise counts can be retrieved from the noise charge spectrum by measuring its total area and the area of the first two peaks, while the rest of the spectrum gives redundant information. A study of the influence of statistical fluctuations is carried out to predict the amount of data needed to achieve a given precision and to study in detail the intrinsic limitations on the possibility to characterize a SiPM where noise is generated according to the model considered. The method is generalizable to other statistical models in order to include devices where noise generation obeys different stochastical processes.


\section{Acknowledgments}
The author would like to thank the valuable support of the IceCube-CTA group at the University of Geneva (Geneva, Switzerland), in particular the many fruitful discussions with Dr.~M.~Heller.


\end{document}